# Cd-substitution effect on photoexcitation properties of ZnO nanodots surrounded by carbon moiety

Ivan Shtepliuk[a,b]

[a] *Semiconductor Materials Division, Department of Physics, Chemistry and Biology-IFM, Linköping University, S-58183 Linköping, Sweden*

[b] *I.M. Frantsevich Institute for Problems of Materials Science, N.A.S. of Ukraine, 3, Krzhizhanovsky Str., UA-03142 Kyiv, Ukraine*

E-mail: ivan.shtepliuk@liu.se

**Abstract**

The geometrical structure and photoexcitation properties of $Zn_{27\text{-}n}Cd_nO_{27}C_{42}$ complexes are investigated by density functional theory (DFT) and time-dependent DFT calculations at the PBE0/6-31G*/SDD level of theory. The cohesive energy and frequency analysis indicate that the hybrid materials are energetically stable. In presence of Cd substituting atoms, energy gap of the ZnO nanodots surrounded by carbon moiety is shown to decrease, as compared to Cd-free complex. In-depth excited state analysis including charge density difference (CDD) mapping and absorption spectrum decomposition is performed to reveal the nature of the dominant excited states and to comprehend the Cd-to-Zn substitution effect on the photoexcitation properties of $Zn_{27\text{-}n}Cd_nO_{27}C_{42}$. A principal possibility to enhance intramolecular charge transfer through incorporation of certain number of Cd atoms into the ZnO nanodots is shown. Such Cd-induced modifications in optical properties of semi-spherical $Zn_{27\text{-}n}Cd_nO_{27}C_{42}$ complexes could potentially enable use of this hybrid material in optoelectronic and photocatalytic applications.

## 1. Introduction

Recently, a theoretical model of novel broadband metamaterial absorber from ultraviolet to near infrared based on a winning combination of zinc oxide and graphene nanodots has been proposed[1]. Interest was focused on the stability and photoexcitation properties of the hybrid materials. Dramatic enhancement of Raman scattering, and a strong light absorption were found for $(ZnO)_{27}C_{42}$ molecule, with the dominating doubly degenerate excited state lying at 635 nm.

The charge density difference (CDD) analysis showed the electron–hole coherence localized at $C_{42}$ carbon moiety, so such this state was characterized to be local excitation (LE). In contrast, high-energy absorption band at 366 nm is mainly related to charge transfer (CT) excited state, implying the obvious electron transfer from the $C_{42}$ moiety to the $(ZnO)_{27}$ fragment. Since intramolecular CT excited states play an important role in the operation of numerous devices like photocatalyzers[2,3,4], solar cells[5,6,7], and light-emitting diodes[8,9,10], their rational manipulation in $(ZnO)_{27}C_{42}$ is critical to improving the material performance, especially towards more efficient photoinduced charge separation. Intuitively, this can be achieved through adjusting the energy between the highest occupied molecular orbital (HOMO) and the lowest unoccupied molecular orbital (LUMO). Indeed, as was demonstrated in the case of conjugated polymers[11], there is a direct causality between HOMO-LUMO energy gap and the intramolecular charge transfer: the narrower the gap, the greater the charge transfer.

The question arises how to reduce the HOMO-LUMO gap of $(ZnO)_{27}C_{42}$ while retaining its beneficial structural and optical properties. A possible solution is an isovalent substitution of Cd for host Zn atoms. This approach has been broadly used to form $Zn_{1-x}Cd_xO$ ternary alloys[12,13,14] possessing narrower band gaps compared to ZnO. Because of the resemblance of the electron affinity and electronegativity of Cd (~0 and 1.69, respectively) with those of Zn (~0 and 1.65, respectively)[15,16] no significant Cd-induced perturbation (at least at low Cd content) of structural properties of ZnO phase is anticipated, except the small lattice volume expansion driven by the difference in the ionic radius[17]. Concomitantly, electronic band structure of ZnO undergoes substantial changes after Cd-to-Zn substitution, manifested as a bang gap shrinkage at the $\Gamma$ point of the first Brillouin zone[18]. The mechanism behind this band-gap narrowing is still questionable and can be associated to several different factors[19] including (i) an upward shift of the valence band maxima (VBM) due to *p*-*d* repulsion arising from the hybridization of O 2*p* state and *d* states of Zn and Cd[20,21], (ii) a downward shift of conduction band maximum (CBM) resulting from the contribution of 5s states of Cd[22] and (iii) Cd-induced tensile strain of ZnO matrix[23].

Back to $(ZnO)_{27}C_{42}$ molecule and bearing in mind the above, one could expect that the energies of its frontier orbitals will be sensitive to the Cd incorporation. Thus, $Zn_{27-n}Cd_nO_{27}C_{42}$ hybrid materials are of significant interest because Cd presence in the matrix of fullerene hemisphere-like nanodots may play a key determining role in intramolecular charge redistribution upon photoexcitation.

The purpose of the present paper is to explore the structural and optical properties of $Zn_{27-n}Cd_nO_{27}C_{42}$ ($n$=1, 2, 3, 4 and 7) hybrid material, with special focus on achieving an understanding of the nature of dominant excited states. Unraveling how Cd substituting atoms affect the photoexcitation properties of $(ZnO)_{27}C_{42}$ will enable more sophisticated design of hybrid materials with enhanced intramolecular charge transfer.

## 2. Methods

The density functional theory (DFT) and time-dependent density functional theory (TD-DFT) calculations reported in this paper were done with Gaussian 16 Rev. C.01 program package[24] under vacuum conditions. $(ZnO)_{27}C_{42}$ molecule passivated with hydrogen atoms was used a basic model for construction of $Zn_{27-n}Cd_nO_{27}C_{42}$ hybrid materials with different number of Cd substituting atoms ($n$=1, 2, 3, 4 and 7, respectively). During geometrical optimization, all atoms were allowed to relax (without any geometrical constraints). SCF (self-consistent field) convergence criterion was set to be $10^{-8}$. The Perdew–Burke–Ernzerhof (PBE0) hybrid exchange-correlation functional that uses 25% exact exchange and 75% DFT exchange was employed in this work[25]. The use of PBE0 functional for investigation of ZnO-based systems is justified by its success at predicting bandgap energy of pure ZnO[26,27,28,29]. Furthermore, since PBE0 also provides an adequate treatment of the electronic properties of graphene derivatives[30,31,32,33,34,35] hybrid exchange-correlation functional is believed to be a compromise choice for the investigation of the photoexcitation properties of $Zn_{27-n}Cd_nO_{27}C_{42}$ system comprising both metal oxide and carbon moiety. 6-31G* basis set[36] has been applied for the carbon, oxygen, and hydrogen atoms, while the SDD basis set developed by the Stuttgart-Dresden-Bonn group[37] has been chosen for zinc and cadmium atoms, respectively. Additional frequency calculations for the optimized structures were performed to ensure the absence of the imaginary frequencies and to confirm finding the true energy minima. The cohesive energy per atom for $Zn_{27-n}Cd_nO_{27}C_{42}$ system was also used as indicative of structural stability[38]:

$$E_{coh} = \frac{E_{Zn_{27-n}Cd_nO_{27}C_{42}} - (\sum_i E_C + \sum_j E_H + \sum_k E_{Zn} + \sum_l E_O + \sum_m E_{Cd})}{n_C + n_H + n_{Zn} + n_O + n_{Cd}} \quad (1)$$

where $E_{Zn_{27-n}Cd_nO_{27}C_{42}}$ is the total energy of the optimized $Zn_{27-n}Cd_nO_{27}C_{42}$ structure, $E_{C,H,Zn,O}$ is the energy of isolated carbon, hydrogen, zinc, oxygen, and cadmium atoms, respectively. $i$, $j$, $k$, $l$ and $m$ indices imply the summation over all C, H, Zn, O and Cd atoms, respectively. $n_{C,H,Zn,O,Cd}$ is the number of corresponding atoms in $Zn_{27-n}Cd_nO_{27}C_{42}$ system.

Photoexcitation properties of the $Zn_{27-n}Cd_nO_{27}C_{42}$ system were investigated by means of TD-DFT calculation at the same level of DFT (PBE0/6-31G*/SDD). The absorption spectra, composed of the lowest 100 excited states, were simulated. The detailed nature of the excited states was further scrutinized by using Multiwfn program[39]. Visualization of both structural and volumetric data was provided by the VESTA program[40].

## 3. Results and discussion

The optimized structures of $Zn_{27-n}Cd_nO_{27}C_{42}$ system with different $n$ values are represented in **Figures 1(a)-(f)** and **Figure S1** (**Supplemental material**). It is clearly seen that the hybrid system preserves its semi-spherical shape, regardless of the number of substituting Cd atoms. Furthermore, $Zn_{27-n}Cd_nO_{27}C_{42}$ becomes more curved with increasing Cd content (**Table 1**). While for all considered structures Zn-O and C-C bonds remain intact, a slight alteration of C-O and C-Zn bond lengths occurs for $Zn_{20}Cd_7O_{27}C_{42}$ structure. One can also conclude that Cd-O and Cd-C bonds are significantly longer than Zn-O and Zn-C counterparts. This is a main reason why the accommodation of Cd atoms in the molecular structure of the hybrid complex is attained through the expansion of $Zn_{27-n}Cd_nO_{27}C_{42}$ matrix that eventually manifests as curving of the structure. **Table 1** also summarizes cohesive energies of $Zn_{27-n}Cd_nO_{27}C_{42}$. Since $E_{coh}$ values of all considered complexes are negative, this entails that there is strong chemical bonding between constituents, indicating the formation of stable structures. Also, it can be deduced from **Table 1** that the Cd-rich complexes are less stable than the $(ZnO)_{27}C_{42}$ complex. In line with the estimation of cohesive energy, the frequency analysis showed that there are no imaginary frequencies, indicating that optimized geometries of $Zn_{27-n}Cd_nO_{27}C_{42}$ correspond to stationary points (global energy minima).

**Table 1** also summarizes the HOMO-LUMO energy gap values of $Zn_{27-n}Cd_nO_{27}C_{42}$. It is obvious that with increasing Cd content HOMO-LUMO energy gap tends to become smaller, reaching the minimum value of ~0.89 eV at $n$=4. The energy gap narrowing is mostly influenced by the downward shift of LUMO level, whereas the HOMO level is less susceptible to the Cd incorporation (**Figure S2, Supplemental material**).

The theoretical Raman spectra of $Zn_{27-n}Cd_nO_{27}C_{42}$ complexes are shown in **Figure 2a**. The first observation is that the fingerprint Raman features of Cd-containing complexes resemble those of $(ZnO)_{27}C_{42}$ complex. The spectra are dominated by three vibrational modes marked as RM1, RM2 and RM3. All three modes are related to the vibrations within carbon moiety. It is worth noting that Cd-induced changes in the chemical bonds are reflected in the shift of the selected modes and the change of corresponding Raman scattering activities (**Figure 2b**). Particularly, it is clearly seen that two most intensive Raman modes, namely RM1 and RM2, demonstrate similar behavior: the bands peaked at 1119.90 and 1384.62 cm$^{-1}$ for Cd-free complex are first red-shifted by 2.4 and 1.1 cm$^{-1}$ for the system with $n$=1 and then shifted upward for the complexes with larger $n$ values. Compared to others, $Zn_{26}Cd_1O_{27}C_{42}$ complex is characterized by the largest Raman scattering activities of both RM1 and RM2 vibrational modes. While a clear attenuation of these Raman modes is observed with Cd content increasing. In contrast, RM3 mode is enhanced in the amplitude and demonstrates downward shift for the Cd-rich $Zn_{20}Cd_7O_{27}C_{42}$ complex. This vibrational mode can be mainly influenced by the asynchronous motion of edge carbon atoms bonded to Zn and Cd atoms.

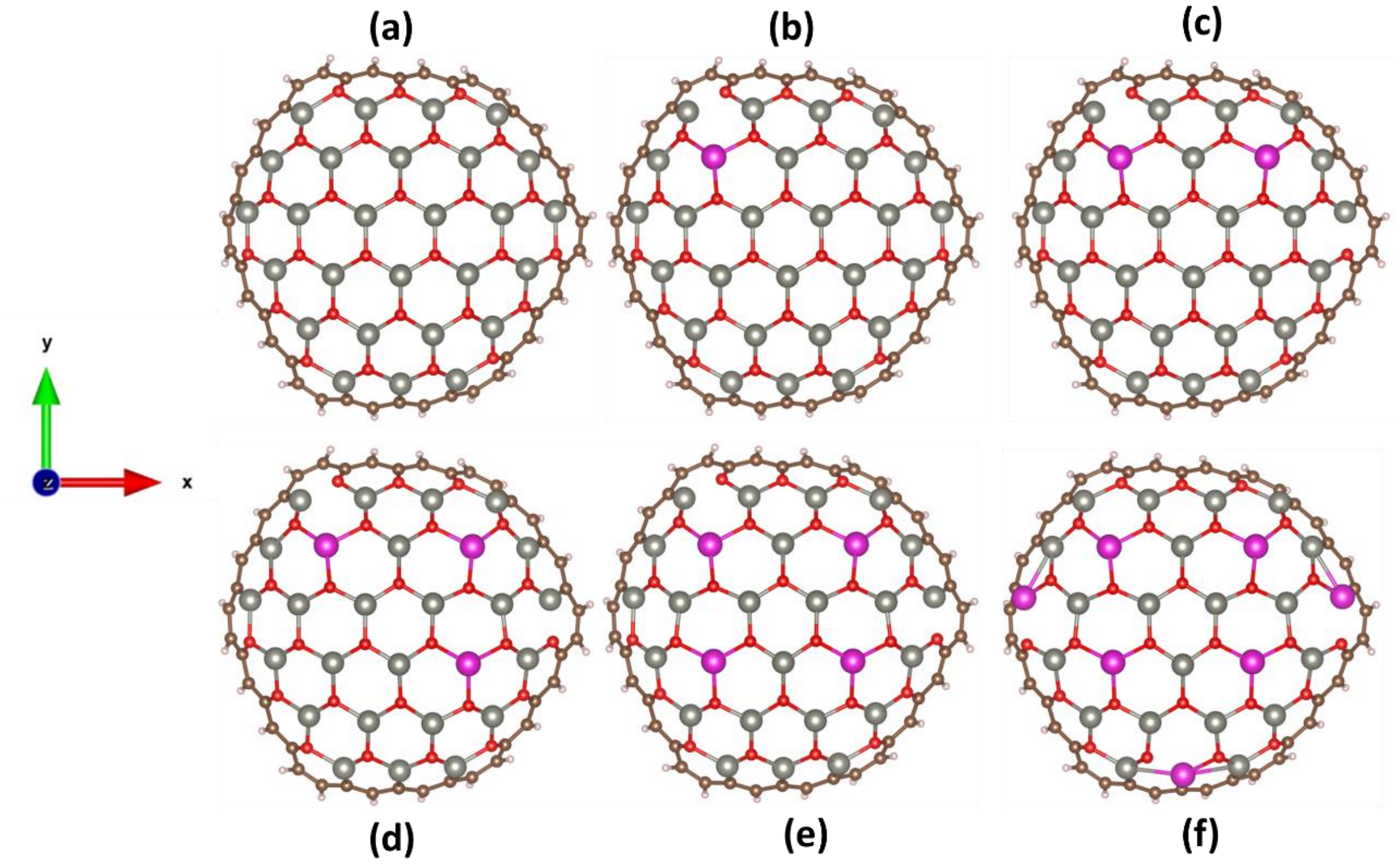

**Figure 1.** (Top view) Optimized structures of the $Zn_{27-n}Cd_nO_{27}C_{42}$: (a) $n$=0, (b) $n$=1, (c) $n$=2, (d) $n$=3, (e) $n$=4 and (f) $n$=7, respectively. Brown, whitish, grey, violet, and red balls correspond to carbon, hydrogen, zinc, cadmium, and oxygen atoms, respectively.

**Table 1.** Cohesive energies, structural parameters, and energy gap values of $Zn_{27-n}Cd_nO_{27}C_{42}$ hybrid structures. The main curvature parameter is determined as the difference between the mean $z$-coordinate and the absolute highest $z$-coordinate in the $Zn_{27-n}Cd_nO_{27}C_{42}$. The difference between minimum and maximum $z$-coordinates is given in square brackets as an additional curvature parameter.

| Number of Cd atoms, *n* | Cohesive energy, eV | Mean Zn-O [Cd-O] bond length, Å | Mean C-C bond length, Å | Mean C-O bond length, Å | Mean C-Zn [C-Cd] bond length, Å | Curvature parameter, Å | Energy gap, eV |
|---|---|---|---|---|---|---|---|
| 0 | -5.8035 | 1.9039 | 1.4091 | 1.3710 | 1.9396 | 3.7695 [6.8327] | 1.2436 |
| 1 | -5.7920 | 1.9050 [2.1059] | 1.4090 | 1.3709 | 1.9398 | 3.8223 [6.9010] | 1.1589 |
| 2 | -5.7805 | 1.9063 [2.1046] | 1.4089 | 1.3709 | 1.9400 | 3.7925 [6.9313] | 1.1151 |
| 3 | -5.7693 | 1.9072 [2.1047] | 1.4088 | 1.3709 | 1.9399 | 4.1047 [7.2986] | 0.9851 |
| 4 | -5.7578 | 1.9084 [2.1035] | 1.4088 | 1.3709 | 1.9397 | 4.1366 [7.3097] | 0.8895 |
| 7 | -5.7257 | 1.9044 [2.1344] | 1.4092 | 1.3673 | 1.9514 [2.1032] | 4.0682 [7.3108] | 0.9644 |

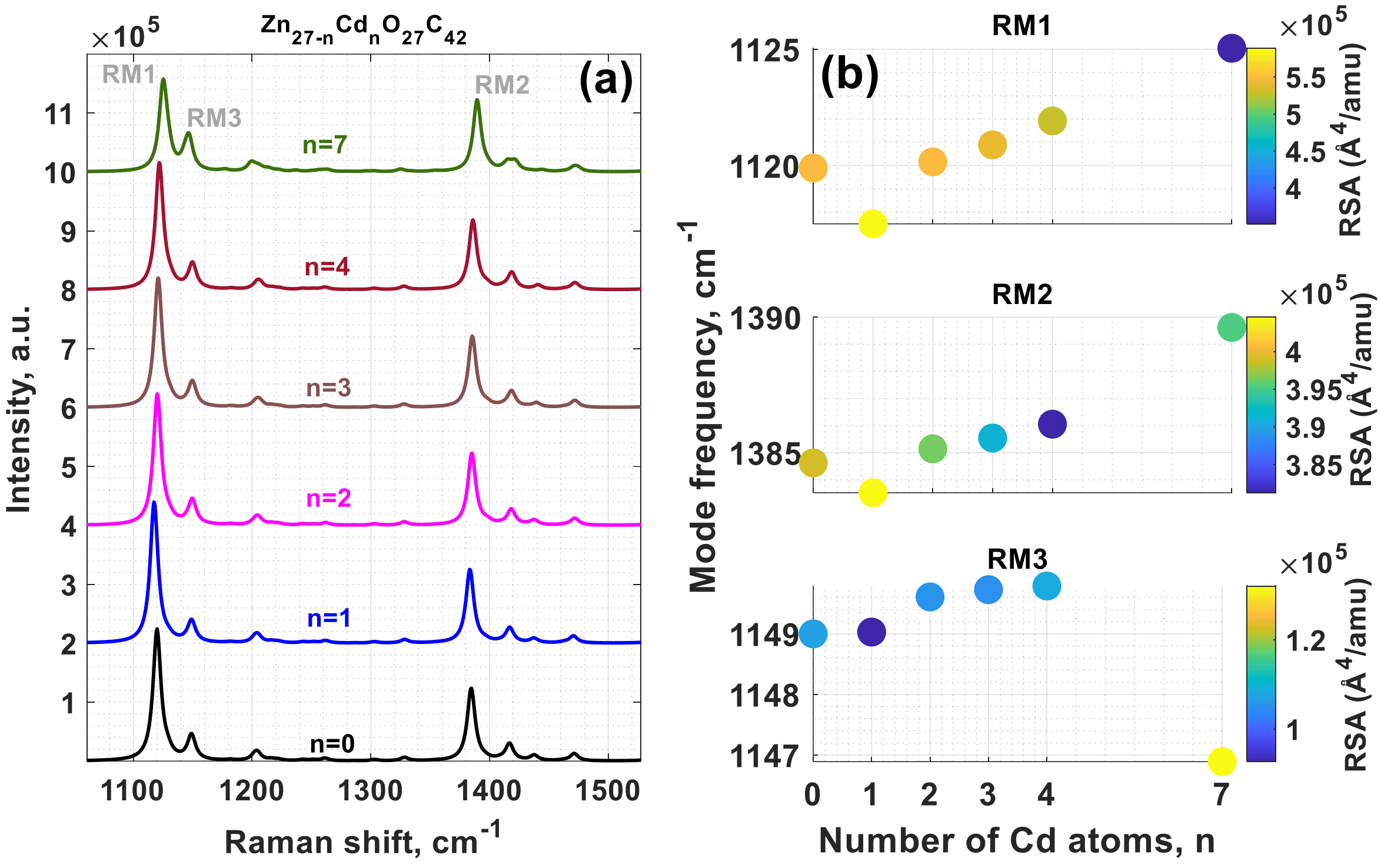

**Figure 2.** (a) Raman spectra of the $Zn_{27\text{-}n}Cd_nO_{27}C_{42}$. (b) Dependences of the frequencies of three selected Raman modes (RM1, RM2 and RM3) on number of Cd atoms color-coded by RSA (Raman scattering activity) values.

The absorption spectra of $Zn_{27\text{-}n}Cd_nO_{27}C_{42}$ complexes are given in **Figure 3**. For the sake of convenience, two different spectral regions are demonstrated separately: 275-450 nm (**Figure 3a**) and 450-700 nm (**Figure 3b**), respectively. The most intensive spectral features in the low-energy side (**Figure 3b**) appearing at about 635 nm have similar shapes for all structures, but this band for Cd-free complex is somewhat higher in the intensity than those of Cd-containing ones. Furthermore, a gradual decrease in the intensity is observed when moving from $n$=1 to $n$=7. It is interesting to note that the spectrum of $(ZnO)_{27}C_{42}$ is mainly dominated by three doubly degenerate excited states: $S_{10,11}$, $S_{23,24}$ and $S_{27,28}$, respectively. Cd-to-Zn substitution breaks the degeneracy as shown in **Table 2**, causing the transition from two-fold degenerate excited states to non-degenerate excites states of different oscillator strengths. **Table S1 (Supplemental material)** also summarizes the properties of remaining excited states with oscillator strengths exceeding 0.1. The most interesting fact related to these absorption spectra is the observation of a diminution of the weak spectral feature (**Figure 3b**) at ~366 nm on the higher-energy side of the main absorption band in Cd-containing complexes with $n$=1 and $n$=2, which continues until the high-energy shoulder completely disappears at $n \geq 3$. Bearing in mind that this absorption band is mainly contributed by CT excited state and the fact that Cd incorporation reduces the HOMO-LUMO gap, it is reasonable to assume that the disappearance of this band is correlated with energy shift of the frontier molecular orbitals. This entails the change of the corresponding CT excited state energies.

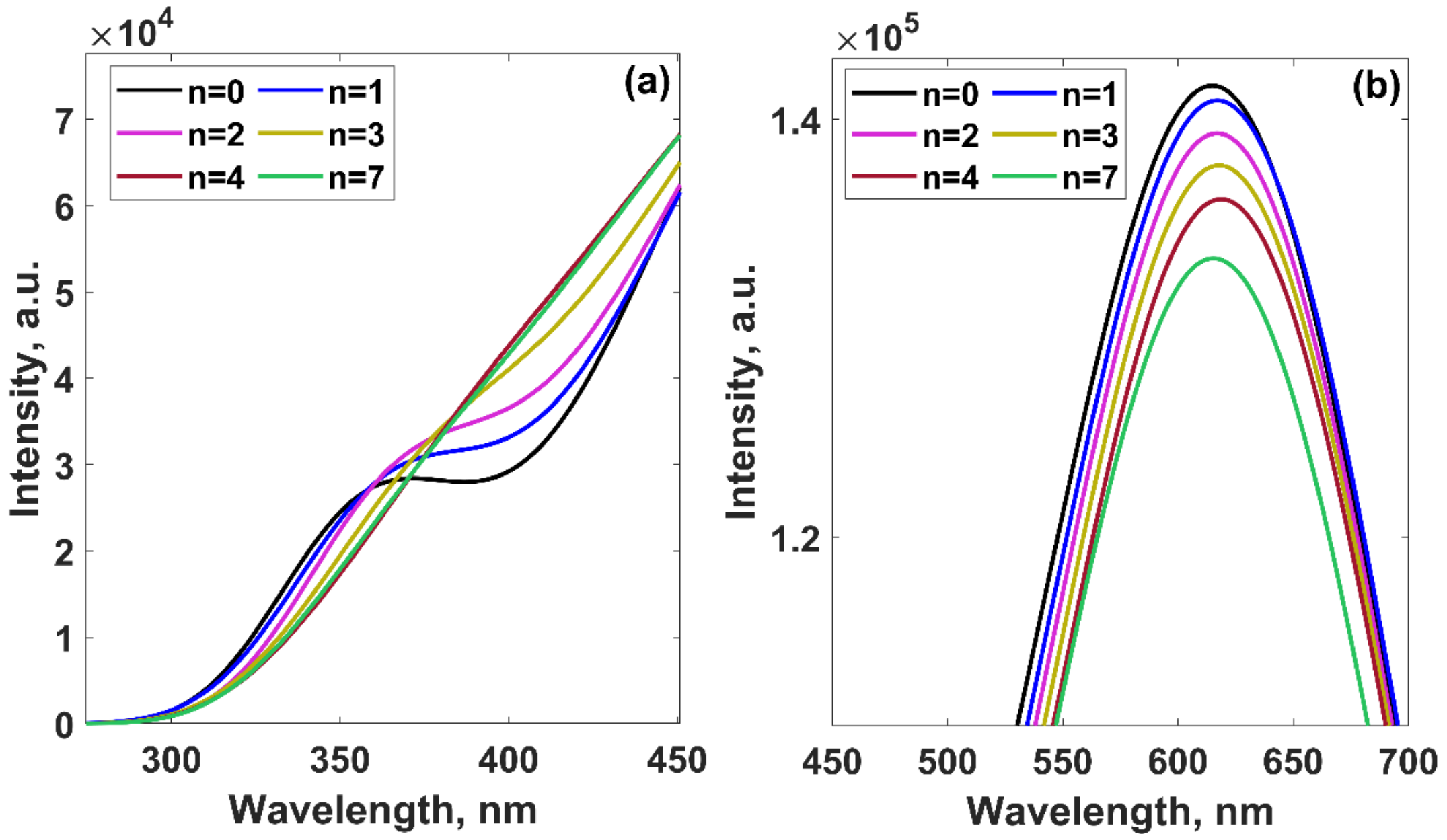

**Figure 3.** (Zoomed views) Absorption spectra of the $Zn_{27-n}Cd_nO_{27}C_{42}$ complexes within different spectral ranges: (a) 275-450 nm and (b) 450-700 nm, respectively**.**

To elucidate the nature of excited states in $Zn_{27-n}Cd_nO_{27}C_{42}$ complexes, the charge density difference (CDD) for selected excited states were then computed (**Figure 4**). It should be mentioned that the yellow-coloured and cyan-coloured CDD regions represent an increase and decrease in the electron density due to the excitation, respectively. As can be seen from **Figure 4a** that at $n$=1 CDD of the excited state of $S_{11}$ and $S_{12}$ is yellow-cyan and mainly located at the carbon moiety, suggesting that these states are locally excited. On the other hand, for CDD of $S_{24}$ state there is an asymmetry in electron density distribution within carbon moiety, which is an indicative of intra-moiety charge transfer. Thus, this state can be classified as a mixed LE-CT state. The same picture was observed for dominating excited states of $Zn_{25}Cd_2O_{27}C_{42}$ system (**Figure 4b**). More insights into the nature of the excited states can be revealed from the analysis of the selected descriptors of electron excitation type[41,42,43] like electron–hole wave-function overlap integral ($S_r$), the distance between centroids of holes and electrons ($D$) and the degree of separation of holes and electrons ($t$). These parameters are listed in **Table 2**. The $D$ parameter for two dominating states in $Zn_{26}Cd_1O_{27}C_{42}$ and $Zn_{25}Cd_2O_{27}C_{42}$ complexes is quite small, while these states are characterized by a large overlap integral and negative *t* parameter. It suggests that no obvious charge transfer occurs, and LE character dominates in these states. In turn, $S_{24}$ and $S_{26}$ states for both structures exhibit increased *D* values (1.18 Å and 1.16 Å, respectively). These results are consistent with results of CDD analysis.

**Table 2**. Properties of dominant excited states in $Zn_{27-n}Cd_nO_{27}C_{42}$ hybrid systems. For doubly degenerate excite states, the properties of second state adjacent to first state are included in square brackets.

| Number of Cd atoms, *n* | Excited state | Wavelength, nm | Oscillator strength, *f* | *D*, Å | $S_r$ | *t*, Å | Type | Major contribution |
|---|---|---|---|---|---|---|---|---|
| 0 | $S_{10,11}$ | 635.52 | 1.55 | 0.24 | 0.89 | -5.25 | LE | H-1->L+4 (46%), HOMO->L+3 (46%) [H-1->L+3 (46%), HOMO->L+4 (46%)] |
| | $S_{23,24}$ | 496.79 | 0.51 | 1.06 | 0.88 | -4.41 | LE-CT | H-2->L+4 (15%), H-1->L+9 (11%), HOMO->L+8 (40%), HOMO->L+9 (19%) [H-2->L+3 (15%), H-1->L+8 (40%), H-1->L+9 (19%), HOMO->L+9 (11%)] |
| | $S_{27,28}$ | 466.6 | 0.27 | 1.33 | 0.86 | -3.63 | LE-CT | H-3->L+3 (19%), H-3->L+4 (10%), H-1->L+8 (12%), HOMO->L+9 (32%) [H-3->L+3 (10%), H-3->L+4 (19%), H-1->L+9 (32%), HOMO->L+8 (12%)] |
| 1 | $S_{11}$ | 638.731 | 1.33 | 0.62 | 0.85 | -1.25 | LE | H-1->L+3 (39%), HOMO->L+4 (41%) |
| | $S_{12}$ | 638.074 | 1.45 | 0.32 | 0.88 | -3.29 | LE | H-1->L+4 (46%), HOMO->L+3 (39%) |
| | $S_{24}$ | 498.368 | 0.48 | 1.18 | 0.87 | -3.97 | LE-CT | H-2->L+4 (16%), H-1->L+9 (28%), HOMO->L+8 (36%) |
| 2 | $S_{13}$ | 636.17 | 1.42 | 0.29 | 0.88 | -2.63 | LE | H-1->L+3 (42%), HOMO->L+4 (43%) |
| | $S_{14}$ | 634.44 | 1.47 | 0.41 | 0.88 | -4.44 | LE | H-1->L+4 (46%), HOMO->L+3 (39%) |
| | $S_{26}$ | 497.88 | 0.44 | 1.16 | 0.85 | -3.52 | LE-CT | H-2->L+3 (12%), H-1->L+8 (46%), HOMO->L+9 (25%) |
| 3 | $S_{13}$ | 640.64 | 0.64 | 2.71 | 0.65 | -0.57 | LE-CT | H-2->L+1 (22%), H-1->L+3 (15%), H-1->L+5 (26%), HOMO->L+4 (14%) |
| | $S_{14}$ | 635.98 | 1.45 | 0.27 | 0.88 | -3.44 | LE | H-1->L+3 (24%), H-1->L+4 (24%), HOMO->L+3 (20%), HOMO->L+4 (26%) |
| | $S_{15}$ | 634.15 | 0.92 | 2.28 | 0.74 | -1.78 | LE-CT | H-1->L+4 (20%), H-1->L+5 (31%), HOMO->L+3 (16%) |
| 4 | $S_{15}$ | 637.35 | 1.44 | 0.37 | 0.88 | -4.32 | LE | H-1->L+3 (30%), H-1->L+4 (17%), HOMO->L+3 (15%), HOMO->L+4 (31%) |
| | $S_{16}$ | 636.40 | 1.49 | 0.13 | 0.89 | -3.82 | LE | H-1->L+3 (16%), H-1->L+4 (34%), HOMO->L+3 (30%), HOMO->L+4 (17%) |
| | $S_{28}$ | 499.47 | 0.45 | 1.21 | 0.86 | -4.01 | LE-CT | H-1->L+8 (20%), HOMO->L+8 (29%), HOMO->L+9 (24%) |
| 7 | $S_{14}$ | 635.84 | 1.04 | 1.32 | 0.81 | -2.93 | LE-CT | H-1->L+4 (42%), HOMO->L+3 (32%), HOMO->L+6 (11%) |
| | $S_{15}$ | 635.29 | 0.66 | 2.20 | 0.66 | 0.09 | CT | H-2->L+1 (14%), H-1->L+3 (22%), H-1->L+5 (22%), HOMO->L+4 (24%), HOMO->L+6 (11%) |
| | $S_{17}$ | 627.22 | 0.60 | 2.83 | 0.61 | -0.58 | LE-CT | H-1->L+3 (14%), HOMO->L+4 (13%), HOMO->L+6 (44%) |

The CDD of $S_{13}$ and $S_{15}$ states demonstrated **Figure 4c** for $Zn_{24}Cd_3O_{27}C_{42}$ shows an obvious charge transfer from carbon moiety to the ZnCdO part. Particularly, it is seen that electrons are mainly accepted by the cadmium atoms and partly accepted by oxygen atoms. In this case, the decrease electron density (cyan-coloured region) is clearly observed at the carbon moiety. Considering the fact that *D* parameters for these states are relatively large (2.71 Å and 2.28 Å) one can say that intra-molecular charge transfer for *n*=3 becomes more pronounced compared to that for complexes with $n$=1 and $n$ =2. However, since $t$ parameter is negative for both states, they cannot be referred to pure CT states. The $S_{14}$ state is locally excited.

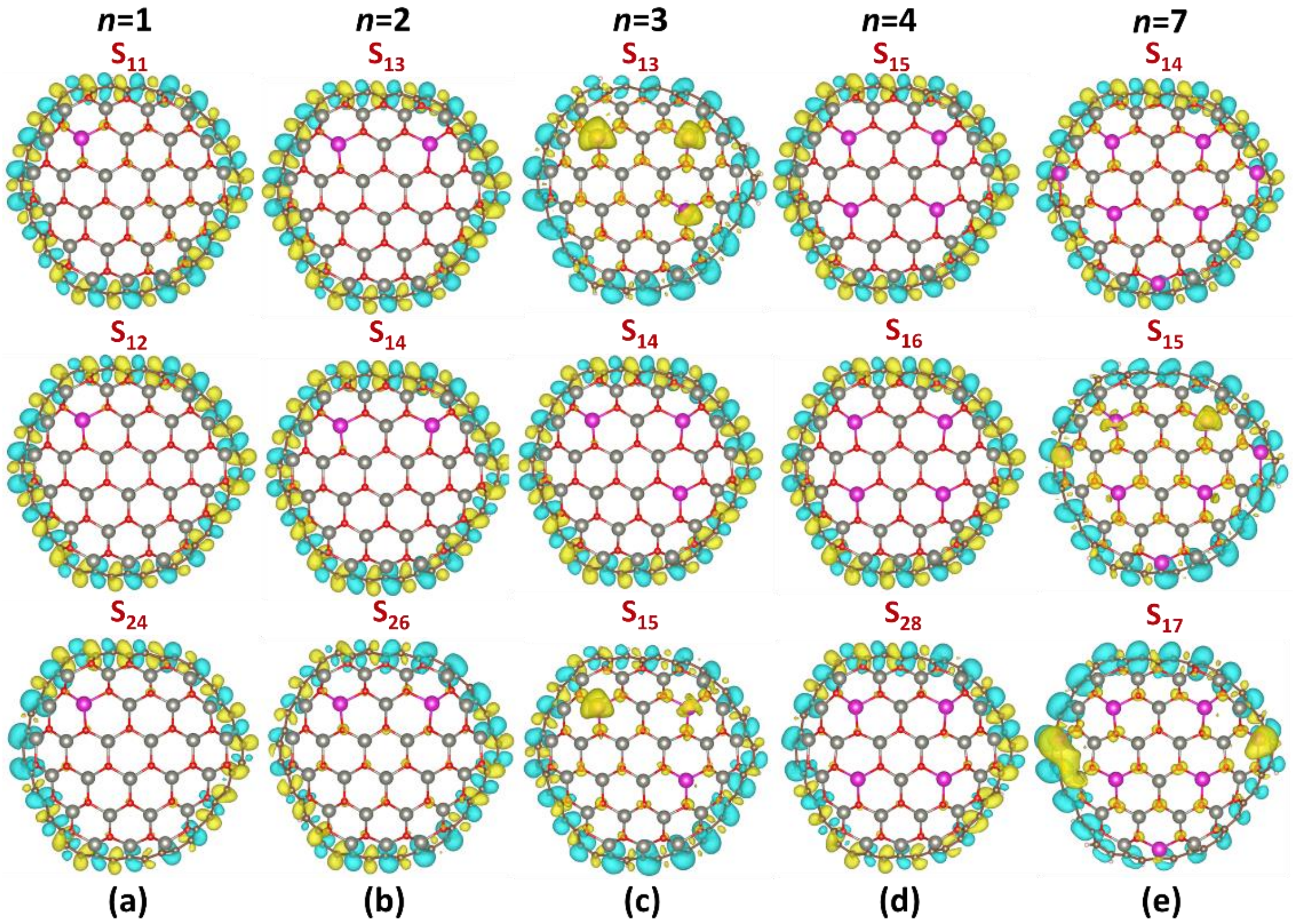

**Figure 4.** CDD for selected excited states (with the largest oscillator strength) in $Zn_{27-n}Cd_nO_{27}C_{42}$ hybrid system: (a) $n$=1, (b) $n$=2, (c) $n$=3, (d) $n$=4 and (e) $n$=7, respectively. Herein, yellow denotes positive charge distribution and cyan means negative charge distribution. CDD was calculated as a difference between the corresponding excited state and the ground state of the considered system ($\rho_{excited} - \rho_{ground}$). Isosurface level is set to be 0.0003.

It is worth mentioning that the incorporation of four Cd atoms in the host molecule makes the molecule more symmetric and hence a more homogeneous charge redistribution is expected. Due to this reason, no decrease of the electron density at both cadmium and oxygen atoms is observed for dominating $S_{15}$, $S_{16}$ and $S_{28}$ states (**Figure 4d**). In contrast, with increasing the amount of Cd atom from $n$=4 to $n$=7, one can observe a significant increase of the electron density on the oxygen and cadmium atoms and the formation of the charge depletion region within the carbon moiety (**Figure 4e**). Furthermore, $S_{15}$ state is characterized by the large $D$ value, positive $t$ value, and the reduced overlap integral, which speaks in favor of an increase of the intra-molecular charge transfer. It is evident that the direction of electron transfer is from the carbon part to the ZnCdO central fragment.

To get a more complete picture of the photoexcitation properties of $Zn_{27-n}Cd_nO_{27}C_{42}$, the charge-transfer spectra (CTS)[44] were plotted and analyzed. To be more specific, CTS concept involves a deconvolution of the absorption spectrum into different components corresponding to intrafragment charge redistribution (ICR) and interfragment charge transfer (ICT). In fact, this approach enables to assess the contribution of charge-transfer excited states to the total absorption spectrum of the complex. The resulting charge-transfer spectra are demonstrated in **Figure 5**.

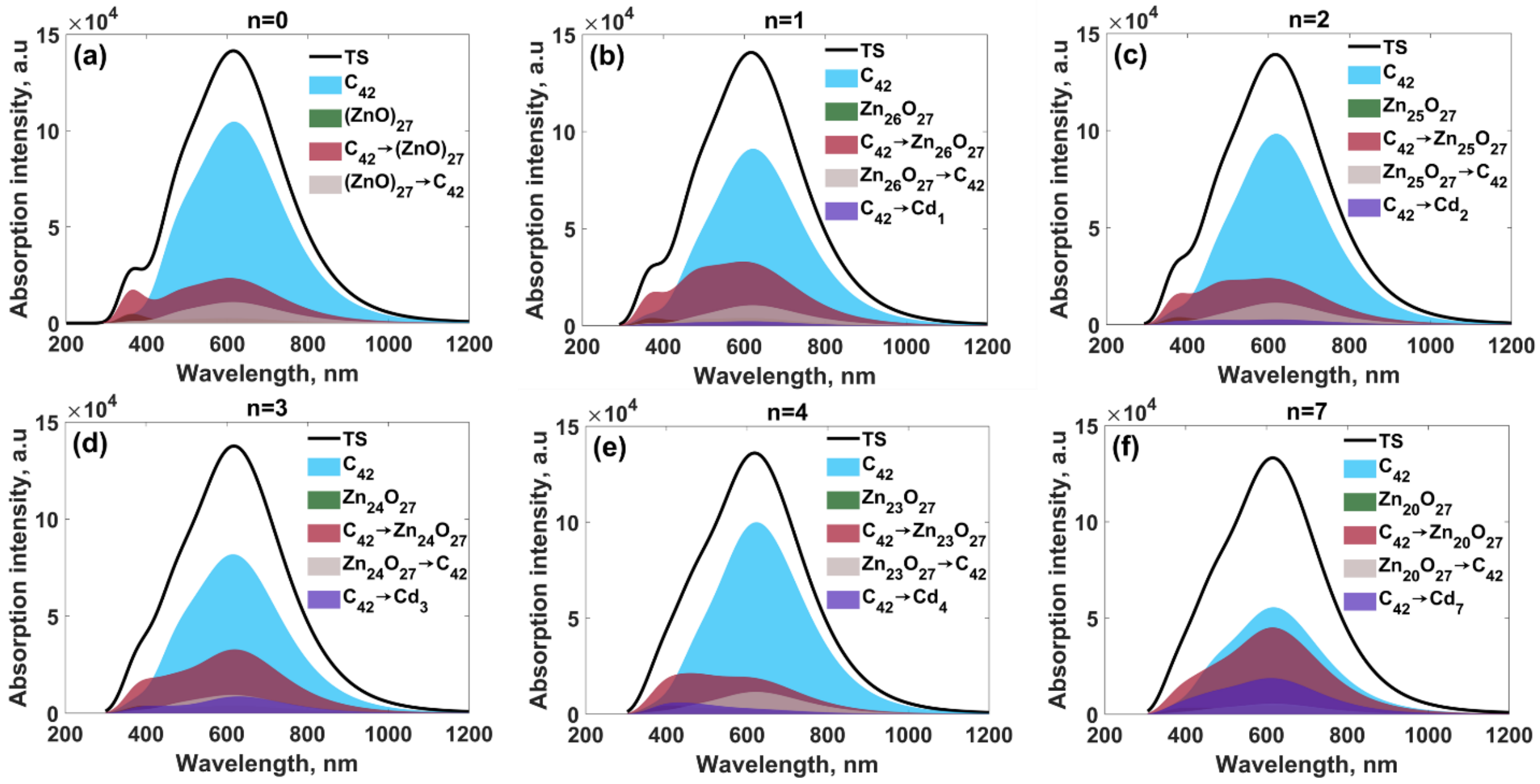

**Figure 5.** Charge transfer spectra of the $Zn_{27-n}Cd_nO_{27}C_{42}$ complexes ranging from 200 nm to 1200 nm: (a) $n$=0, (b) $n$=1, (c) $n$=2, (d) $n$=3, (e) $n$=4 and (f) $n$=7, respectively. Black solid lines denoted as TS correspond to the total absorption spectra of the considered systems. Horizontal arrow in the figure legend indicates the direction of the charge transfer between selected fragments.

The analysis of CTS evolution shows that the dominant feature of the absorption spectra of hybrid complexes with $n$ ranging from 0 up to 4 is caused by the electron transitions within the $C_{42}$ moiety. However, an equitable contribution from the ICR and ICT components to the CTS is observed at $n$=7, additionally highlighting a significant role of the substituting Cd atoms in the charge-transfer processes due to the photoexcitation. Notably, in all cases a prevailing ICT component is attributed to the charge transfer from $C_{42}$ to ZnO fragment. Although the component that is associated with the charge transfer from carbon moiety to the cadmium atoms can be also distinguished. Furthermore, an increase of the contribution from this component to the total spectrum is observed when reaching the $n$ value of 7. A final observation is that the structures where ZnO nanodots doped with odd number of Cd atoms exhibit stronger ICT components than those with even number of Cd atoms. This points to the principal possibility to tune the intramolecular CT excited states not through Cd content control alone, but by implementing parity-odd effects.

## 4. Conclusions

In conclusion, the structural and optical properties of novel-type $Zn_{27-n}Cd_nO_{27}C_{42}$ complexes were studied by first principle calculations with PBE0 functional. The stability of the hybrid complexes was confirmed by estimating the cohesive energy and performing the frequency analysis. Cd-to-Zn substitution was found to decrease the HOMO-LUMO energy gap, mainly due to the shift of the LUMO level. A Cd-induced switch between local excited states and charge transfer states in ZnO nanodots surrounded by carbon moiety was revealed and ascribed to the appearance of the effective interfragment charge transfer from the carbon moiety to the ZnCdO part. It was also found that the complexes with odd number of Cd atoms exhibit improved intramolecular charge transfer compared to those with even number of substituting Cd atoms, probably due to the non-uniform electron density distribution. This provides a useful tool to realize the ZnO-based broadband absorber material by adjusting Cd:Zn atomic ratio for the next-generation applications in photocatalysis and optoelectronics. The obtained results create prerequisites for more-sophisticated experimental design of metal oxide-carbon-based materials with desired light absorption ability.

## Acknowledgments

Author acknowledges the support from Ångpanneföreningens Forskningsstiftelse (Grant 21–112). The computations and data handling were enabled by resources provided by the National Supercomputer Centre (NSC), funded by Linköping University. This work was also supported by the project "Creation of effective photocatalytic materials based on ZnO-Cd system for water purification from organic pollutants", funded by NAS of Ukraine.

## Disclosure statement

The authors report there are no competing interests to declare.

## References

[1]I. Shtepliuk, R. Yakimova, Nature of photoexcited states in ZnO-embedded graphene quantum dots, preprint, arxiV, (2022). link

[2]C. Zhu, T. Wei, Y. Wei, L. Wang, M. Lu, Y. Yuan, L. Yin and L. Huang, J. Mater. Chem. A **9**, 1207-1212 (2021). doi: 10.1039/D0TA08609F

[3] W. Zhao, D.Zhai, C. Liu, D. Zheng, H. Wu, L.Sun, Z. Li, T. Yu, W. Zhou, X. Fang, S. Zhai, K. Han, Z. He, W.-Q. Deng, Appl. Catal., B **300**, 120719 (2022). doi: 10.1016/j.apcatb.2021.120719

[4] S. Wan, J.Xu, S. Cao, J. Yu, Interdiscip. Mater. **1**, 294–308 (2022). doi: 10.1002/idm2.12024

[5] R. Qin, W.Yang, S.Li, T.-K.Lau, Z.Yu, Z. Liu, M. Shi, X. Lu, C.-Z. Li, and H. Chen, Mater. Chem. Front. **3**, 513-519 (2019). doi: 10.1039/C8QM00609A

[6] K. Chordiya, Md. Ehesan Ali, and M. U. Kahaly, ACS Omega **7**, 16, 13465–13474 (2022). doi: 10.1021/acsomega.1c06233

[7] L. Fu, H. Hu, Q.Zhu, L. Zheng, Y.Gu, Y.Wen, H. Ma, H. Yin, J. Ma, Nano Res. (2022). doi: 10.1007/s12274-022-5000-4.

[8] X. Ouyang, X.-L. Li, X. Zhang, A. Islam, Z. Ge, S.-J. Su, Dyes Pigm. **122**, 264-271 (2015).doi: 10.1016/j.dyepig.2015.06.036

[9] M. Bayda-Smykaj, K. Rachuta, G. L. Hug, M. Majchrzak, and B. Marciniak, J. Phys. Chem. C **125**, 23, 12488–12495 (2021). doi: 10.1021/acs.jpcc.1c00990.

[10] H. Tanaka, K. Shizu, H. Nakanotani, C. Adachi, Chem. Mater. **25**, 18, 3766–3771 (2013). doi: 10.1021/cm402428a.

[11] D. Hashemi, X. Ma, R. Ansari, J. Kim and J. Kieffer, Phys. Chem. Chem. Phys. **21**, 789-799 (2019). doi: 10.1039/C8CP03341B.

[12] I. Shtepliuk, V. Khranovskyy, G. Lashkarev, V. Khomyak, A. Ievtushenko, V. Tkach, V. Lazorenko, I. Timofeeva, R. Yakimova, Appl. Surf. Sci. **276**, 550-557 (2013). doi: 10.1016/j.apsusc.2013.03.132.

[13] I. Shtepliuk, G. Lashkarev, V. Khomyak, O. Lytvyn, P. Marianchuk, I. Timofeeva, A. Ievtushenko, V. Lazorenko, Thin Solid Films, **520**, 14, 4772-4777 (2012). doi: 10.1016/j.tsf.2011.10.181.

[14] Y. Z. Zhu, G. D. Chen, Honggang Ye, Aron Walsh, C. Y. Moon, Su-Huai Wei, Phys. Rev. B **77**, 245209 (2008). doi: 10.1103/PhysRevB.77.245209.

[15] L. Pauling, *The Nature of the Chemical Bond*, 3rd ed. (Cornell University Press, United States, 1960) p. 93.

[16] T. R. Myers, J. Chem. Educ. **67**, 307 (1990). doi: 10.1021/ed067p307.

[17] R. D. Shannon, Acta Crystallogr. A. **32** (5), 751–767 (1976). doi: 10.1107/S0567739476001551.

[18] M. Gandouzi, Z.R.Khan, A. S. Alshammari, Comput. Mater. Sci. **156**, 346-353 (2019). doi: 10.1016/j.commatsci.2018.09.056.

[19] X. Tang, H.-F. Lu, J.-J. Zhao, Q.-Y. Zhang, J. Phys. Chem. Solids **71**, 3, 336-339 (2010). doi: 10.1016/j.jpcs.2009.12.086.

[20] A. Schleife, F. Fuchs, J. Furthmüller, F. Bechstedt, Phys. Rev. B **73**, 245212 (2006). doi: 10.1103/PhysRevB.73.245212.

[21] S.H. Wei, A. Zunger, Phys. Rev. B 37, 8958 (1988). doi: 10.1103/PhysRevB.37.8958.

[22] X. D. Zhang, M. L. Guo, W. X. Li, and C. L. Liu, J. Appl. Phys. **103**, 063721 (2008). doi: 10.1063/1.2901033.

[23] S. Li-Bin, K.Li, J. Jian-Wei and C. Feng, Chinese Phys. B **18**, 4418 (2009). doi: 10.1088/1674-1056/18/10/052.

[24] M. J. Frisch, G. W. Trucks, H. B. Schlegel, G. E. Scuseria, M. A. Robb, J. R. Cheeseman, G. Scalmani, V. Barone, G. A. Petersson, H. Nakatsuji, X. Li, M. Caricato, A. V. Marenich, J. Bloino, B. G. Janesko, R. Gomperts, B. Mennucci, H. P. Hratchian, J. V. Ortiz, A. F. Izmaylov, J. L. Sonnenberg, D. Williams-Young, F. Ding, F. Lipparini, F. Egidi, J. Goings, B. Peng, A. Petrone, T. Henderson, D. Ranasinghe, V. G. Zakrzewski, J. Gao, N. Rega, G. Zheng, W. Liang, M. Hada, M. Ehara, K. Toyota, R. Fukuda, J. Hasegawa, M. Ishida, T. Nakajima, Y. Honda, O. Kitao, H. Nakai, T. Vreven, K. Throssell, J. A. Montgomery, Jr., J. E. Peralta, F. Ogliaro, M. J. Bearpark, J. J. Heyd, E. N. Brothers, K. N. Kudin, V. N. Staroverov, T. A. Keith, R. Kobayashi, J. Normand, K. Raghavachari, A. P. Rendell, J. C. Burant, S. S. Iyengar, J. Tomasi, M. Cossi, J. M. Millam, M. Klene, C. Adamo, R. Cammi, J. W. Ochterski, R. L. Martin, K. Morokuma, O. Farkas, J. B. Foresman, and D. J. Fox, Gaussian 16, Revision C.01 (Gaussian, Inc., Wallingford CT, 2016).

[25] C. Adamo, V. Barone, J. Chem. Phys. **110**, 6158-69 (1999). doi: 10.1063/1.478522.

[26] A. Chesnokov, D. Gryaznov, N. V. Skorodumova, E.A. Kotomin, A. Zitolo, M. Zubkins, A. Kuzmin, A. Anspoks and J. Purans, J. Mater. Chem. C **9**, 4948-4960 (2021). doi: 10.1039/D1TC00223F.

[27] F. Viñes, F. Illas, J. Comput. Chem. **38**, 523–529 (2017). doi: 10.1002/jcc.24705.

[28] L. I. Bendavid, L. H. Kugelmass, Surf. Sci. **695**, 121575 (2020). doi: 10.1016/j.susc.2020.121575.

[29] M. Gerosa, C. E. Bottani, L. Caramella, G. Onida, C. Di Valentin, and G. Pacchioni, Phys. Rev. *B* **91**, 155201 (2015). doi: 10.1103/PhysRevB.91.155201.

[30] S. M. Elgengehi, S. El-Taher, M. A. A. Ibrahim, J. K. Desmarais, K. E. El-Kelany, Appl. Surf. Sci. **507**, 145038 (2020). doi: 10.1016/j.apsusc.2019.145038.

[31] J. Paier, M. Marsman, G. J. Kresse, Chem. Phys. **127**, 024103 (2007). doi: 10.1063/1.2747249

[32] Z. Qian, J. Ma, X. Shan, L. Shao, J. Zhou, J. Chen, H. Feng, RSC Adv. **3**, 14571– 14579 (2013). doi: 10.1039/C3RA42066C.

[33] A. Ramasubramaniam, Phys. Rev. B: Condens. Matter Mater. Phys. **81**, 245413 (2010). doi: 10.1103/PhysRevB.81.245413.

[34] M. Zhao, F. Yang, Y. Xue, D. Xiao, Y. A. Guo, ChemPhysChem **15**, 950–957(2014). doi: 10.1002/cphc.201301137.

[35] I. Shtepliuk, R. Yakimova, PCCP **20** (33), 21528-21543 (2018). doi: 10.1039/C8CP03306D.

[36] R. Krishnan, J. S. Binkley, R. Seeger, J. A. J. Pople, Chem. Phys. **72**, 650–654 (1980). doi: 10.1063/1.438955.

[37] J. M. L. Martin, A. Sundermann, J. Chem. Phys. **114**, 3408-3420 (2001). doi: 10.1063/1.1337864.

[38] I. Shtepliuk, V. Khranovskyy, R. Yakimova, PCCP **19** (45), 30445-30463 (2017). doi: 10.1039/C7CP04711H

[39] T. Lu, F. Chen, J. Comput. Chem. **33**, 580–592 (2012). doi: 10.1002/jcc.22885.

[40] K. Momma, F. Izumi, J. Appl. Crystallogr. **44**, 1272–1276 (2011). doi: 10.1107/S0021889811038970.

[41] B. Moore, H. Sun, N. Govind, K. Kowalski and J. Autschbach, J. Chem. Theory Comput. **11**, 3305-3320 (2015). doi: 10.1021/acs.jctc.5b00335.

[42] W. Ding, Z. Xue, J. Li, M. Li, L. Bai, Q. Zhou, X. Zhou, Y. Peng, and L. Miao, ECS J. Solid State Sci. Technol. **11**, 016001 (2022). doi: 10.1149/2162-8777/ac4c80.

[43] I. Shtepliuk, R. Yakimova, Materials **11** (7), 1217 (2018). doi: 10.3390/ma11071217.

[44] Z. Liu, X. Wang, T. Lu, A.Yuan, X. Yan, Carbon **187**, 78-85 (2022). doi: 10.1016/j.carbon.2021.11.005.

# Supplemental material
# for
# Cd-substitution effect on photoexcitation properties of ZnO-embedded graphene quantum dots

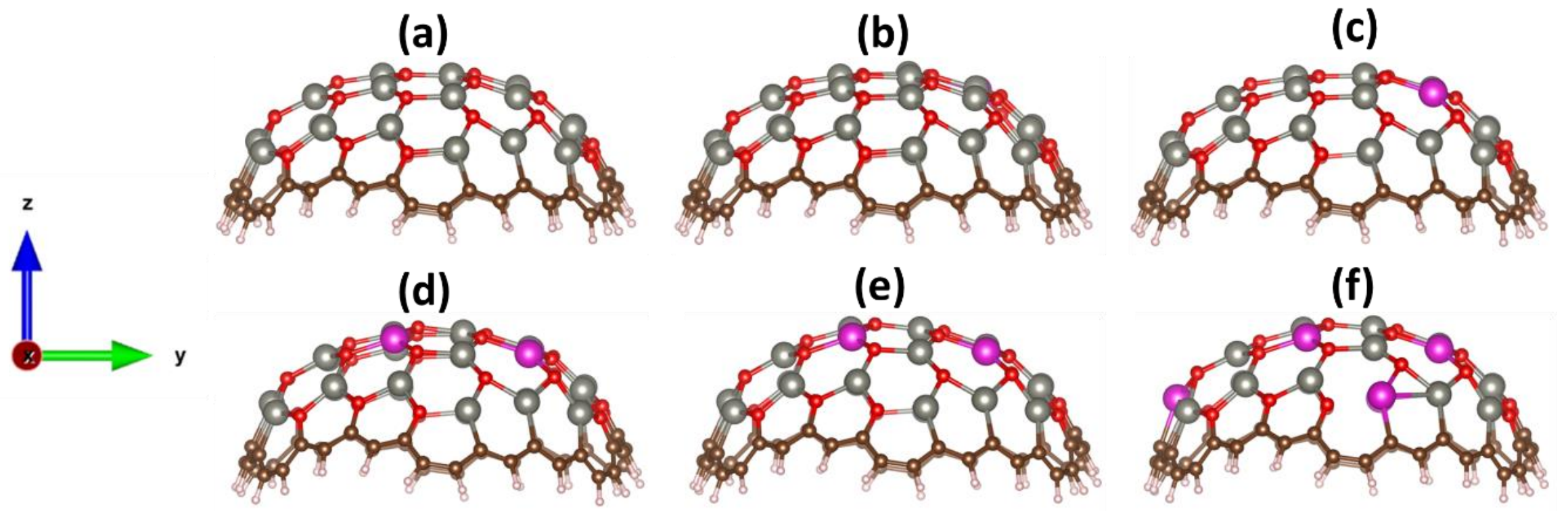

**Figure S1.** (Side view) Optimized structures of the $Zn_{27-n}Cd_nO_{27}C_{42}$: (a) $n$=0, (b) $n$=1, (c) $n$=2, (d) $n$=3, (e) $n$=4 and (f) $n$=7, respectively. Brown, whitish, grey, violet, and red balls correspond to carbon, hydrogen, zinc, cadmium and oxygen atoms, respectively.

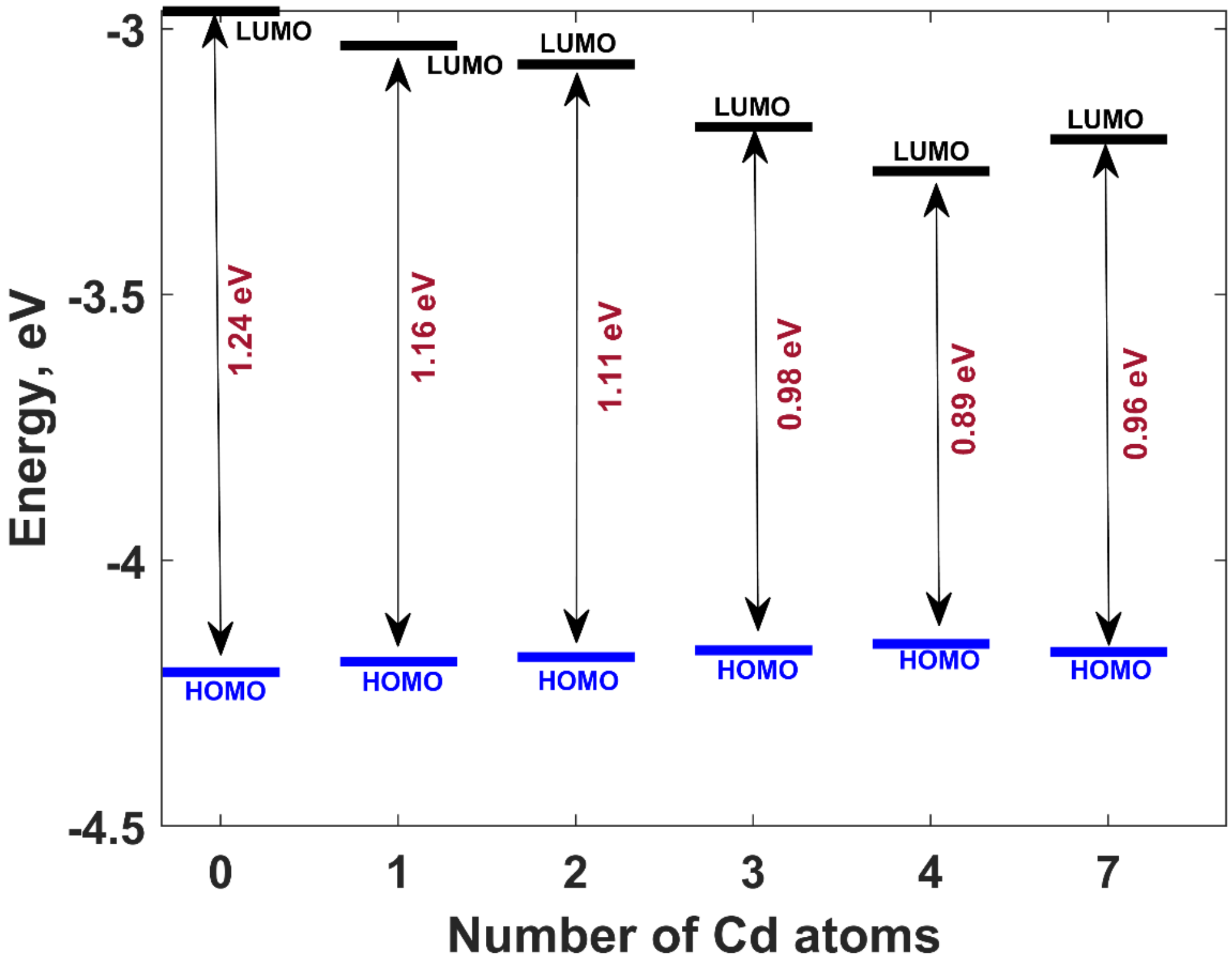

**Figure S2.** Molecular orbital diagram summarizing all considered systems. The arrows designate HOMO–LUMO gaps.

**Table S1**. Properties of excited states with *f* > 0.1 in $Zn_{27-n}Cd_nO_{27}C_{42}$ hybrid systems.

| Number of Cd atoms, *n* | Excited state | Wavelength, nm | Oscillator strength, *f* | *D*, Å | $S_r$ | *t*, Å | Major contribution |
|---|---|---|---|---|---|---|---|
| 0 | $S_{19}$ | 535.45 | 0.10 | 1.52 | 0.79 | -3.86 | H-2->L+3 (68%) |
| | $S_{20}$ | 535.43 | 0.10 | 1.52 | 0.79 | -3.91 | H-2->L+4 (68%) |
| 1 | $S_{13}$ | 630.38 | 0.26 | 2.99 | 0.49 | 1.025 | H-2->L+1 (52%), H-1->L+5 (19%) |
| | $S_{19}$ | 538.05 | 0.12 | 1.36 | 0.88 | -4.01 | H-2->L+3 (70%) |
| | $S_{23}$ | 502.81 | 0.11 | 3.02 | 0.51 | 1.36 | H-5->LUMO (21%), H-3->L+1 (40%), HOMO->L+7 (19%) |
| | $S_{25}$ | 497.86 | 0.39 | 1.44 | 0.81 | -2.37 | H-2->L+3 (12%), H-1->L+8 (34%), HOMO->L+9 (29%) |
| | $S_{29}$ | 468.39 | 0.16 | 2.23 | 0.65 | 0.44 | H-2->L+5 (18%), H-1->L+9 (10%), H-1->L+10 (30%), HOMO->L+8 (12%) |
| | $S_{30}$ | 466.31 | 0.17 | 1.84 | 0.74 | -0.90 | H-3->L+2 (10%), H-3->L+3 (31%), H-1->L+8 (18%), HOMO->L+9 (18%) |
| | $S_{31}$ | 465.86 | 0.11 | 1.71 | 0.76 | -1.04 | H-3->L+4 (29%), H-2->L+5 (15%), H-1->L+9 (22%) |
| | $S_{62}$ | 374.50 | 0.13 | 2.69 | 0.61 | -0.43 | H-11->LUMO (15%),H-10->LUMO (14%),H-4->L+3 (10%),H-1->L+17 (11%) |
| 2 | $S_{12}$ | 655.58 | 0.11 | 3.76 | 0.37 | 1.12 | H-2->L+1 (21%), H-1->L+5 (53%) |
| | $S_{19}$ | 538.33 | 0.13 | 1.36 | 0.89 | -4.23 | H-2->L+3 (70%) |
| | $S_{25}$ | 499.27 | 0.44 | 1.16 | 0.85 | -3.66 | H-2->L+4 (15%), H-1->L+9 (22%), HOMO->L+8 (46%) |
| | $S_{31}$ | 468.41 | 0.18 | 1.53 | 0.83 | -3.01 | H-3->L+3 (33%), H-1->L+8 (12%), HOMO->L+9 (31%) |
| | $S_{32}$ | 465.72 | 0.27 | 1.16 | 0.83 | -2.63 | H-3->L+4 (27%),H-1->L+9 (35%), HOMO->L+8(10%),HOMO->L+10 (12%) |
| | $S_{62}$ | 378.46 | 0.13 | 2.46 | 0.63 | -0.20 | H-11->LUMO (24%), H-1->L+17 (22%) |
| 3 | $S_{20}$ | 538.73 | 0.11 | 1.12 | 0.87 | -4.25 | H-2->L+3 (64%) |
| | $S_{26}$ | 500.07 | 0.43 | 1.16 | 0.85 | -3.67 | H-2->L+4 (16%), H-1->L+9 (20%), HOMO->L+8 (45%) |
| | $S_{27}$ | 498.93 | 0.45 | 1.19 | 0.85 | -3.73 | H-2->L+3 (13%), H-1->L+8 (46%), HOMO->L+9 (24%) |
| | $S_{31}$ | 468.97 | 0.21 | 1.48 | 0.85 | -3.49 | H-3->L+3 (33%), H-1->L+8 (15%), HOMO->L+9 (37%) |
| | $S_{32}$ | 466.72 | 0.24 | 1.25 | 0.86 | -3.56 | H-3->L+4 (30%), H-1->L+9 (40%), HOMO->L+8 (13%) |
| | $S_{60}$ | 389.11 | 0.11 | 3.59 | 0.53 | 0.23 | H-11->LUMO (42%) |
| 4 | $S_{23}$ | 537.54 | 0.12 | 1.35 | 0.84 | -3.42 | H-2->L+3 (61%) |
| | $S_{27}$ | 503.83 | 0.15 | 3.03 | 0.54 | 0.36 | H-2->L+5 (57%) |
| | $S_{29}$ | 496.51 | 0.33 | 1.61 | 0.81 | -2.67 | H-2->L+5 (21%), H-1->L+8 (24%), H-1->L+9 (15%), HOMO->L+8 (17%) |
| | $S_{31}$ | 469.26 | 0.18 | 1.51 | 0.84 | -3.22 | H-3->L+3 (28%), H-1->L+8 (15%), HOMO->L+9 (37%) |
| | $S_{32}$ | 467.37 | 0.22 | 1.30 | 0.86 | -3.49 | H-3->L+4 (27%), H-1->L+9 (39%), HOMO->L+8 (14%) |
| | $S_{48}$ | 417.60 | 0.10 | 2.49 | 0.63 | 0.17 | H-11->LUMO (16%), H-3->L+3 (14%) |
| 7 | $S_{13}$ | 645.38 | 0.21 | 3.15 | 0.46 | 1.27 | H-1->L+5 (34%), H-1->L+6 (22%), HOMO->L+6 (15%) |
| | $S_{16}$ | 629.16 | 0.39 | 3.04 | 0.55 | 0.26 | H-1->L+3 (10%),H-1->L+5(19%), H-1->L+6 (32%),HOMO->L+4 (12%) |
| | $S_{27}$ | 508.09 | 0.10 | 3.39 | 0.50 | 0.41 | H-2->L+5 (16%), H-1->L+8 (42%), HOMO->L+8 (14%) |
| | $S_{28}$ | 500.38 | 0.12 | 2.63 | 0.60 | 0.76 | H-3->L+2 (32%), H-1->L+8 (13%) |
| | $S_{29}$ | 496.51 | 0.11 | 3.21 | 0.46 | 1.70 | H-3->L+2 (42%), HOMO->L+9 (13%) |
| | $S_{30}$ | 491.94 | 0.45 | 1.08 | 0.82 | -1.39 | H-1->L+9 (32%), H-1->L+10 (13%), HOMO->L+10 (29%) |
| | $S_{31}$ | 488.68 | 0.15 | 2.09 | 0.70 | -0.57 | H-2->L+5 (29%), H-1->L+10 (12%), HOMO->L+11 (14%) |
| | $S_{34}$ | 466.35 | 0.12 | 1.32 | 0.82 | -1.69 | H-3->L+3 (23%), H-2->L+6 (14%), H-1->L+9 (22%), HOMO->L+10 (20%) |
| | $S_{35}$ | 460.18 | 0.12 | 2.17 | 0.78 | -2.40 | H-3->L+4 (24%), HOMO->L+9 (13%), HOMO->L+11 (32%) |
| | $S_{36}$ | 457.57 | 0.13 | 2.53 | 0.66 | -0.79 | H-3->L+3 (13%), H-2->L+6 (20%), H-1->L+11 (38%) |